&&&&&&&&&&&&&&&&&&&&&&&&&&&&&&&&&&&&&&&&&&&&&&&&&&&&&&&&&&&&&&&&&&&&&&&&&
\documentstyle[12pt]{article}

\def\thefootnote{\fnsymbol{footnote}}
\def\bea{\begin{eqnarray}}
\def\eea{\end{eqnarray}}
\def\beq{\begin{equation}}
\def\eeq{\end{equation}}
\def\W{\overline{W}}
\def\cW{{\cal W}}
\def\cbW{{\cal \W}}

\def\tF{{\tilde F}}

\def\hcA{\hat{{\cal A}}}

\def\hA{\hat{A}}

\def\notcD{\not{\hspace{-.05in}{\cal D}}}

\def\notD{\not{\hspace{-.05in}D}}

\def\lll{{\ln\Lambda^2\over 32\pi^2}}

\def\ba{\bar{\alpha}}

\def\[{\left [}
\def\]{\right ]}
\def\({\left (}
\def\){\right )}
\def\lbr{\left\{}
\def\rbr{\right\}}
\def\pp{\partial}
\def\M{\bar{M}}

\def\y{\bar{y}}
\def\z{\bar{z}}

\def\STr{{\rm STr}}
\def\Tr{{\rm Tr}}
\def\cA{{\cal A}}
\def\K{{\cal K}}

\def\f{\bar{f}}
\def\F{{\cal F}}

\def\L{{\cal L}}
\def\D{{\cal D}}
\def\bl{\bar{\lambda}}
\def\hf{\hat{\phi}}
\def\hz{\hat{z}}

\def\hA{\hat{A}}

\def\hD{\hat{D}}

\def\hV{\hat{V}}
\def\n{\bar{n}}
\def\m{\bar{m}}
\def\s{\bar{s}}
\def\cM{{\cal{M}}}

\def\bc{\bar{\chi}}

\def\A{\bar{A}}
\def\bc{\bar{\chi}}
\def\bps{\bar{\psi}}

\def\Z{{\bar{Z}}}

\def\bF{\bar{F}}

\begin{document}

\begin{titlepage}
\begin{center}

\hfill LBL-37697 \\
\hfill UCB-PTH-95/31 \\
\hfill ITP-SB-95-38 \\ 
\hfill hep-th/9606135 \\
\hfill \today \\

\vskip .3in
{\large \bf SUPERGRAVITY COUPLED TO CHIRAL AND YANG-MILLS MATTER
AT ONE LOOP}\footnote{This 
work was supported in part by
the Director, Office of Energy Research, Office of High Energy and Nuclear
Physics, Division of High Energy Physics of the U.S. Department of Energy under
Contract DE-AC03-76SF00098 and in part by the National Science Foundation under
grants PHY-95-14797 and PHY-93-09888.}

Mary K. Gaillard$^{\em a}$, Vidyut Jain${\em ^b}$\footnote{Present address: 
vid@dotcomdev.com} {\em and} Kamran Saririan$^{\em a}$

{\em $^a$Physics Department and Theoretical Physics Group,
 Lawrence Berkeley Laboratory, 
 University of California, Berkeley, California 94720}\\

{\em $^b$Institute for Theoretical Physics, SUNY at Stony Brook, NY 11794}

\end{center}

\begin{abstract}

We present the full result for the divergent one-loop contribution to the 
effective boson Lagrangian for supergravity coupled to chiral and Yang-Mills
supermultiplets.  We also consider the specific case of dilaton couplings in
effective supergravity Lagrangians from superstrings, for which the one-loop
result is considerably simplified.

\end{abstract}
\end{titlepage}
\newpage

\renewcommand{\thepage}{\roman{page}}
\setcounter{page}{2}
\mbox{ }

\vskip 1in

\begin{center}
{\bf Disclaimer}
\end{center}

\vskip .2in

\begin{scriptsize}
\begin{quotation}
This document was prepared as an account of work sponsored by the United
States Government. While this document is believed to contain correct 
 information, neither the United States Government nor any agency
thereof, nor The Regents of the University of California, nor any of their
employees, makes any warranty, express or implied, or assumes any legal
liability or responsibility for the accuracy, completeness, or usefulness
of any information, apparatus, product, or process disclosed, or represents
that its use would not infringe privately owned rights.  Reference herein
to any specific commercial products process, or service by its trade name,
trademark, manufacturer, or otherwise, does not necessarily constitute or
imply its endorsement, recommendation, or favoring by the United States
Government or any agency thereof, or The Regents of the University of
California.  The views and opinions of authors expressed herein do not
necessarily state or reflect those of the United States Government or any
agency thereof, or The Regents of the University of California.
\end{quotation}
\end{scriptsize}

\vskip 2in

\begin{center}
\begin{small}
{\it Lawrence Berkeley Laboratory is an equal opportunity employer.}
\end{small}
\end{center}

\newpage
\renewcommand{\thepage}{\arabic{page}}
\setcounter{page}{1}
\def\thefootnote{\arabic{footnote}}
\setcounter{footnote}{0}

In this Letter we present the ultraviolet divergent contributions~\cite{us,us2}
to the one-loop corrections for a
general supergravity theory~\cite{crem,bggm}, with only the restriction that 
the gauge kinetic function is diagonal in gauge indices: $f_{ab}(Z) = 
\delta_{ab}k_af(Z)$.  The details of the calculations are given 
in~\cite{us,us2}.  Here we give only sufficient details 
to establish notation and to specify the gauge fixing and our prescriptions for
the loop expansion of the effective action. 

The tree-level supergravity Lagrangian~\cite{crem,bggm} we adopt, with 
$f(z) = x + iy$ (which is trivially generalized~\cite{us2} to $f_{ab} = 
\delta_{ab}k_af,\;k_a=$ constant, and so includes all known string models), is  
\bea {1\over\sqrt{g}}\L &=& {1\over 2}r - {x\over 4}F_{\mu\nu}F^{\mu\nu} - 
{y\over 4}{\tilde F}_{\mu\nu}F^{\mu\nu} + K_{i\m}\D^\mu z^i\D_\mu\z^{\m}- V 
\nonumber \\ & & + {ix\over 2}\bl\notD\lambda + 
iK^{i\m}\(\bc_L^{\m}\notD\chi_L^i + \bc_R^i\notD\chi_R^{\m}\) \nonumber \\ & & 
+ e^{-K/2}\({1\over 4}f_i\A^i\bl_R\lambda_L - A_{ij}\bc^i_R\chi^j_L
+ {\rm h.c.}\)  \nonumber \\ & & +
{1\over 4}\bps_\mu\gamma^\nu(i\notD + M)\gamma^\mu\psi_\nu 
- {1\over 4}\bps_\mu\gamma^\mu(i\notD + M)\gamma^\nu\psi_\nu \nonumber \\ & &
-\[{x\over 8}\bps_\mu\sigma^{\nu\rho}\gamma^\mu\lambda_aF^a_{\nu\rho}
+ \bps_\mu\notcD \z^{\m}K_{i\m}\gamma^\mu L\chi^i 
{1\over 4}\bps_\mu\gamma^\mu\gamma_5\lambda^a\D_a 
+ i\bps_\mu\gamma^\mu L\chi^im_i + {\rm h.c.}\]
\nonumber \\ & & + \(i\bl^a_R\[2K_{i\m}(T_a\z)^{\m} - 
{1\over 2x}f_i\D_a - {1\over 4}\sigma_{\mu\nu}F^{\mu\nu}_af_i\]\chi^i_L + 
{\rm h.c.}\) \nonumber \\ & & + {\rm 4\;fermion \;terms}, \eea
where 
\bea V &=& \hV + \D, \;\;\;\; \hV = e^{-K}(A_i\A^i - 3A\A), \quad
\D = {1\over 2x}\D_a\D^a, \;\;\;\; \D_a = K_i(T^az^i),  \nonumber \\ 
\M &=& (M)^{\dag} = e^{K/2}\(WR + \W L\), \nonumber \\ A &=& e^KW, \;\;\;\; 
\A = e^K\W, \;\;\;\; m_i = e^{-K/2}A_i. \eea
$K(z,\z)$ is the K\"ahler potential, $W(z)$ is the superpotential, $T^a$ is a
generator of the gauge group, and
\beq A_{i_1\cdots i_n} = D_{i_1}\cdots D_{i_n}A, \;\;\;\; \A^{i_1\cdots i_n} = 
D^{i_1}\cdots D^{i_n}\A, \qquad D^i = K^{i\m}D_{\m},\eeq
with $D_i$ the scalar field reparameterization covariant derivative, and
$K^{i\m}$ the inverse K\"ahler metric.
The one-loop effective action is determined from the quadratic quantum action:
\bea &&\L_{quad}(\Phi,\Theta,c) = {1\over 2}\hf^I\hf^J\(\pp_I\pp_J + 
(A_I)^K_J\pp_K\)S + \L_{gf} + \L_{gh} =
\nonumber \\ && \qquad -{1\over 2}\Phi^TZ^\Phi\(\hD^2_\Phi + H_\Phi\)\Phi 
+ {1\over 2}\bar{\Theta}Z^\Theta\(i\notD_\Theta - M_\Theta\)\Theta \nonumber 
\\ & & \qquad + {1\over 2}\bar{c} Z^c\(\hD^2_c + H_c\)c + O(\psi_{cl}), \eea
where $\phi^I = \Phi^I,\;\Theta^I,\; \pp_I = {\pp/\pp\phi^I},$
and the column vectors,
$$\Phi^T = (h_{\mu\nu},\hcA^a,\hz^i,\hz^{\m}),\;\;\;\;\;
\Theta^T = (\psi_\mu,\lambda^a,\chi^i_L,\chi^{\m}_R,\alpha),\;\;\;
\; c^T = (c_\nu, c^a, c_\alpha),$$
represent the boson, fermion and ghost quantum degrees of freedom, 
respectively, with
$ \alpha = -C\ba^T$ an auxiliary field introduced~\cite{us} to implement 
gravitino gauge fixing.  The connection $(A_I)^K_J$ in (4), which is defined
explicitly in~\cite{us,us2}, is chosen so as to preserve
all bosonic symmetries, and also to simplify matrix elements involving the
graviton. In particular the quantum variables 
$\hz^i,\hz^{\m}$ are normal coordinates in the space of scalar fields: 
$(A_i)^k_j = \Gamma^k_{ij}$ is the affine connection associated with the 
K\"ahler metric $K_{i\m}$, giving a scalar field reparameterization invariant 
expansion.  In (4) $\psi_{cl}$ represents background fermion fields that we 
set to zero; that is, we calculate only the one-loop bosonic action.

For the boson sector, we use a smeared gauge-fixing:
\bea \L &\to& \L + \L_{gf}, \;\;\;\; 
\L_{gf} = -{\sqrt{g}\over 2}C_AZ^{AB}C_B,\;\;\;\; \nonumber \\ 
Z &=& \pmatrix{\delta^{ab}&0\cr0& -g^{\mu\nu}\cr}, \;\;\;\; 
C = \pmatrix{C_a\cr C_\mu\cr}. \eea
The Yang-Mills gauge-fixing term:
$$ C^a = \(\D_\mu + {\pp_\mu x\over2x}\)\hcA^a_\mu + 
{i\over\sqrt{x}}K_{i\m}\[(T^a\z)^{\m}\hz^i - (T^az)^i\hz^{\m}\],$$
preserves off-shell supersymmetry~\cite{barb} in the limit of global
supersymmetry and coincides with the string-loop result~\cite{ant} for chiral
multiplet wave function renormalization. The graviton gauge-fixing term:
$$ \sqrt{2}C_\mu = \(\nabla^\nu h_{\mu\nu}
- {1\over 2}\nabla_\mu h^\nu_\nu - 2\D_\mu z^IZ_{IJ}\hz^J + 
2\F^a_{\mu\nu}\hcA_a^\nu \), $$
is the one originally introduced by 't Hooft and Veltman~\cite{tini},
generalized~\cite{us} to include the Yang-Mills sector.
The script quantum and classical Yang-Mills fields and field strengths are
canonically normalized~\cite{noncan}:
$$ \cA_\mu = \sqrt{x}A_\mu, \;\;\;\; \hcA_\mu = \sqrt{x}\hA_\mu, \;\;\;\;
\F_{\mu\nu} = \sqrt{x}F_{\mu\nu}, \;\;\;\; 
\sqrt{x}\D_\mu A_\nu = \(\D_\mu - {\pp_\mu x\over2x}\)\cA_\mu,$$ 
where $\D_\mu$ is the gauge and general covariant derivative. 
In the earlier literature two gravitino gauge fixing procedures have been used:
a) a Landau-type gauge~\cite{ino,josh} $\gamma\cdot\psi = 0$, implemented by
the introduction of an auxiliary field, and b) a smeared gauge~\cite{lahanas}
$\L\to\L - \bF\cM F,\; F = \gamma\cdot\psi, \; \cM = {1\over 4}\(i\notD 
+ 2M\)$ supplemented with Nielsen-Kallosh ghosts.  Here we use an unsmeared
gauge $G = 0,$ with the gauge-fixing function~\cite{us} 
\bea
G &=& -\gamma^\nu(i\notD - \M)\psi_\nu 
- 2(\notcD z^iK_{i\m}R\chi^{\m} + \notcD\z^{\m}K_{i\m}L\chi^i) \nonumber \\ & &
+ {x\over 2}\sigma^{\nu\rho}\lambda_aF^a_{\nu\rho} 
+ 2im_I\chi^I - \gamma_5\D_a\lambda^a, \eea
where $D_\mu$ contains the spin and chiral K\"ahler connections.
The quantum Lagrangian is obtained by the introduction of an auxiliary field
$\alpha$: $\delta(G) = \int d\alpha\;{\rm exp}\(i\alpha G\), $
and a shift in the gravitino field:
$ \psi' = \psi +\gamma\alpha, \;\bps' = \bps + \ba\gamma, $
so as to diagonalize the gravitino kinetic energy term. 
The ghost and ghostino determinants are obtained in the usual way as,
respectively:
$$ \(\hD^2_c + H_c\)^A_B = {\pp\over\pp\epsilon_A}\delta C_B, 
\;\;\;\; A,B = a,\mu,\;\;\;\;
\(\hD^2_c + H_c\)^\alpha_\beta = {\pp\delta G^\alpha\over\pp\epsilon^\beta}, $$
where $\hD_\mu$ is related to $\D_\mu$ or $D_\mu$ by additional
connections~\cite{us,us2}.
With these choices the one-loop bosonic action takes a very simple form: 
\beq S_1 = {i\over 2}\Tr\ln\(\hD_\Phi^2 + H_\Phi\) 
-{i\over 2}\Tr\ln\(-i\notD_\Theta + M_\Theta\) 
+ {i\over2}\STr\ln\(\hD_c^2 + H_c\), \eeq
where
\bea \STr\ln\(\hD_c^2 + H_c\) &=& 2\Tr\ln\(\hD_c^2 + H_c\)_{c_\alpha} 
- 2\Tr\ln\(\hD_c^2 + H_c\)_{c_{a,\mu}}, \nonumber \eea
which just reduces to determinants of the form of those for
scalars and spin-${1\over2}$ fermions. Moreover the ghost and nonghost
sectors have separately supersymmetric quantum spectra, except for the
Yang-Mills fields:
$$ {1\over 2}\(\Tr \;1\)_\Theta = \(\Tr \;1\)_\Phi -2N_G =  2N+2N_G+10.\;\;\;\;
\(\Tr \;1\)_{c_\alpha} = 4, \;\;\;\; \(\Tr \;1\)_{c_{a,b}} = 4 + N_G, $$
where $N (N_G)$ is the number of chiral (gauge) supermultiplets.
To evaluate (7) we separate~\cite{josh,us2} the fermion determinant into
helicity-even and -odd parts:
\bea
-{i\over 2}\Tr\ln(-i\notD + M_\Theta) \equiv -{i\over 2}\Tr\ln\cM(\gamma_5)
 = T_- + T_+, \eea
where here $D_\mu$ contains all fermion connections, and 
\bea T_- &=& -{i\over 4}\[\Tr\ln\cM(\gamma_5) - \Tr\ln\cM(-\gamma_5)\],\qquad
\nonumber \\ 
T_+ &=& -{i\over 4}\[\Tr\ln\cM(\gamma_5) + \Tr\ln\cM(-\gamma_5)\], \nonumber \\
\cM(\gamma_5) &=&\gamma_0(-i\notD + M_\Theta) = \pmatrix{\sigma_+^\mu D^+_\mu 
& M^+\cr M^- &\sigma_-^\mu D^-_\mu\cr},\quad
\sigma_{\pm}^\mu = (1, \pm\vec \sigma). \eea
Then defining
$$ \hD^2_\Theta + H_\Theta \equiv
\(-i\notD_\Theta + M_\Theta\)\(i\notD_\Theta + M_\Theta\), $$ 
The one-loop bosonic action (7) reduces to: 
\bea S_1 &=& {i\over 2}\STr\ln\(\hD^2 + H\) + T_-. \nonumber \eea
The helicity-odd term $T_-$ is at most logarithmically divergent, and is 
finite~\cite{us2} in the absence of a dilaton, that is, for 
$f(Z) = g^{-2} + i\theta/8\pi^2 =$ constant.  As discussed in~\cite{mk,us2}
there is an ambiguity in the separation (9) of the fermion determinant into 
helicity-even and -odd contributions, because terms that are even and odd in 
$\gamma_5$ can be interchanged by the use of the identities:
\beq \gamma_5  = (i/24)\epsilon^{\mu\nu\rho\sigma}\gamma_\mu
\gamma_\nu\gamma_\rho\gamma_\sigma, \;\;\;\;
\sigma_{\mu\nu} = i\gamma_5\sigma^{\rho\sigma}\epsilon_{\rho\sigma\mu\nu}, 
\;\;\;\; etc.\eeq
In most cases the choice is dictated by gauge or K\"ahler covariance.  However 
supersymmetry must be used to fix the off-diagonal gaugino-$\alpha$ and
gaugino-dilaton mass terms:
\beq M_{\alpha\lambda^a} = -\sqrt{x\over2}F_a^{\mu\nu}\sigma_{\mu\nu},
\;\;\;\; M_{\chi^i\lambda^a} = -i{f_i\over4\sqrt{x}}\( F_a^{\mu\nu} 
- i \gamma_5\tF_a^{\mu\nu}\)\sigma_{\mu\nu},\eeq
and the dilaton-dependent gaugino connection: \bea
A^\mu_{\lambda^a\lambda^b} &=& \delta_{ab}{\pp^\mu y\over2x} 
{\epsilon^{\lambda\nu\rho\sigma}\over24}\gamma_\lambda
\gamma_\nu\gamma_\rho\gamma_\sigma.\eea 
Eqs. (11) and (12) are precisely the choices that allow Pauli-Villars 
regularization
of the quadratic divergences~\cite{mk}.  The choice (12) further insures the
nonrenormalization~\cite{russ} of the topological charge $\theta = 8\pi^2y$, 
and is consistent with linear-chiral multiplet duality~\cite{frad} for the
dilaton supermultiplet.

With the above conventions we obtain~\cite{us,us2} the following result for the
divergent contributions to the one-loop corrected
effective supergravity Lagrangian:
\bea \L_{eff} &=& \L(K) + \L_{1-loop} = 
\L(K_R) + \sqrt{g}{\Lambda^2\over32\pi^2}L_0 + \sqrt{g}
\lll\(L_1 + L_2 + N_GL_g\) ,\nonumber \\ 
L_0 &=& {r\over2}\(N_G -1\) + 2K_{i\m}\(\D_\mu z^i\D^\mu\z^{\m} + 5F^i\bF^{\m}\)
- 8\D + {28\over9}M\M + 2x^{-1}\D_aD_i(T^az)^i \nonumber \\ & & +
2N\(K_{i\m}F^i\bF^{\m}- {2\over9}M\M - {r\over4}\) - \(N_G{f_i\f_{\m}\over2x^2} 
+ 2R_{i\m}\)\(F^i\bF^{\m} + \D_\mu z^i\D^\mu\z^{\m}\) , \nonumber \\ 
L_1 &=& {N + 5\over 27}\[K_{i\m}\(2F^i\bF^{\m} - \D_\mu z^j\D^\mu\z^{\n}\)M\M
-{1\over3}(M\M)^2\] \nonumber \\ & & 
+ {N + 5\over 3}K_{i\m}K_{j\n}\(\D_\mu z^j\D^\mu\z^{\m}F^j\bF^{\n} +
\D_\mu z^j\D^\mu z^i\D_\nu\z^{\m}\D^\nu\z^{\n}\) \nonumber \\ & & 
+ 8\(K_{j\n}F^j\bF^{\n} -{1\over3}M\M\)\[K_{i\m}\(F^i\bF^{\m} + 
\D_\mu z^j\D^\mu\z^{\n}\)-{1\over3}M\M\] \nonumber \\ & & 
- 4\(\D_\mu\z^{\m}\D^\mu z^iK_{i\m}\)^2 - 
8\D_\mu z^j\D^\mu z^i\D_\nu\z^{\m}\D^\nu\z^{\n}K_{i\n}K_{j\m} \nonumber \\ & & 
- \(\D_\mu z^i\D^\mu\z^{\m} + F^i\bF^{\m}\)\[{2\over 3}K_{i\m}R_{j\n}
\(\D_\mu z^j\D^\mu\z^{\n} + F^j\bF^{\n}\) + 4R_{\n i\m j}\bF^{\n}F^j \]
\nonumber \\ & & + {1\over3}\D_\mu z^i\D_\nu\z^{\m}K_{i\m}\(R_{j\n}-2K_{j\n}\)
\(\D^\mu z^j\D^\nu\z^{\n} - \D^\nu z^j\D^\mu\z^{\n}\)\nonumber \\ & & 
- \lbr\(\D_\mu z^i\D^\mu\z^{\m} + F^i\bF^{\m}\)\[e^{-K}R^{j\;k}_{\;n\;i}
A_{jk}\A^n_{\m} + e^{-K/2}(D_{\m}R^{j\;k}_{\;n\;i})A_{jk}F^n\] + {\rm h.c.}\rbr
\nonumber \\ & & 
- e^{-K/2}\bigg[{1\over3}R^{k\;\ell}_{\;j\; i}A_{k\ell}\(\M\D_\mu z^i\D^\mu z^j 
+ MF^iF^j\) \nonumber \\ & & \qquad
+ \D_\mu z^i\D^\mu z^jF^m\(R^{k\;\ell}_{\;j\; m}A_{ik\ell} - 
R^{k\;\ell}_{\;j\; i}A_{mk\ell}\) + {\rm h.c.}\bigg] \nonumber \\ & & 
+ R^{k\;\;\ell}_{\;\;j\;\; i}R_{\n k\m\ell}\(\D_\mu z^j\D^\mu z^i
\D_\nu\z^{\n}\D^\nu\z^{\m} - F^jF^i\bF^{\n}\bF^{\m}\) \nonumber \\ & & 
+ R^k_{j\m i}R^i_{\ell\n k}\(\D_\mu z^j\D^\mu\z^{\m}\D_\nu z^{\ell}\D^\nu\z^{\n}
+ 2\D_\mu z^j\D^\mu\z^{\m}F^{\ell}\bF^{\n} + F^j\bF^{\m}F^{\ell}\bF^{\n}\)
\nonumber \\ & & 
+ \D_\mu z^j\D_\nu\z^{\m}R^k_{i\m j}
\D^\mu z^{\ell}\D^\nu\z^{\n}R^i_{k\n \ell}
- \D_\mu z^j\D_\nu\z^{\m}R^k_{i\m j}
\D^\nu z^{\ell}\D^\mu\z^{\n}R^i_{k\n \ell},\nonumber \\ 
L_2 &=& \[\cW^{ab}\(3C_G\delta_{ab} - D_i(T_bz)^jD_j(T_az)^i\) + 
{\rm h.c.}\] + 2M\M\D + 22\D^2 
\nonumber \\ & &  + {N + 5\over6}\[\(\cW^{ab} + \cbW^{ab}\)\D_a\D_b -
x\(F^a_{\rho\mu} -i\tF^a_{\rho\mu}\)
\(F_a^{\rho\nu} + i\tF_a^{\rho\nu}\)\D_\nu z^i\D^\mu\z^{\m}K_{i\m}\]
\nonumber \\ & &  + {N+ 5\over3}\[x^2\cW_{ab}\cbW^{ab} + 2\D^2 - \D
K_{i\m}\(\D_\rho z^i\D^\rho\z^{\m} + 2F^i\bF^{\m}\)
+ {2\over9}\D M\M\] 
\nonumber \\ & & + 12\(\cW^{ab} + \cbW^{ab}\)\D_a\D_b 
+ x\(\cW + \cbW\)\[{2\over3}M\M + K_{i\m}\(
\D_\rho z^i\D^\rho\z^{\m} - 2F^i\bF^{\m}\)\] \nonumber \\ & &  
-2x\(\cW +\cbW\)\D + 2\D K_{i\m}\(8\D_\rho z^i\D^\rho\z^{\m} + 11F^i\bF^{\m}\)
- 26i\D_\mu z^j\D_\nu\z^{\m}K_{i\m}\D^aF_a^{\mu\nu} \nonumber \\ & &  
+ 14x^2\cW_{ab}\cbW^{ab} + \(\D_\mu z^i\D^\mu\z^{\m} + F^i\bF^{\m}\)\[{2\over x}
R_{\n i\m j}\D_a D^j(T^a\z)^{\n}  + {4\over3}\D R_{i\m}\] \nonumber \\ & &  
+ {D_i(T_az)^i\over6x}\lbr4\D_a\[\(\D_\mu z^j\D^\mu\z^{\m} + F^j\bF^{\m}\)
K_{j\m} - 2\D\] + 13iF^a_{\mu\nu}K_{\m j}\D^\mu z^j\D^\nu\z^{\m}\]
\nonumber \\ & & 
+ 2iF^a_{\mu\nu}D_j(T_az)^iR^j_{i\m k} \D^\mu z^k\D^\nu\z^{\m} -
{e^{-K/2}\over x}\D_a\[(T^az)^iR_{i\;\;\ell}^{\;\; j\;\; k}F^{\ell}A_{jk} + 
{\rm h.c.}\] \nonumber \\ & &  + {f_i\f_{\m}\over2x^2}\(F^i\bF^{\m} + 19
\D_\mu z^i\D^\mu\z^{\m}\)\D - \D\(4 + {f_k\f^k\over6x^2}\)
\({f_i\over x}F^iM + {\rm h.c.}\) \nonumber \\ & &  
+ {\f_{\m}f_j\over2x}\bF^{\m}F^j\(\cW + \cbW\) + 
{f_k\f^k\f_{\m}f_j\over8x^4}\D\(\D_\mu\z^{\m}\D^\mu z^j - \bF^{\m}F^j\)
\nonumber \\ & &  - {f_i\f^{\m}\over x}\[{1\over4}\(F^a_{\mu\nu} + 
i\tF^a_{\mu\nu}\)\(F_a^{\mu\rho} - i\tF_a^{\mu\rho}\)\D_\rho z^i\D^\nu\z^{\m} 
+ {11i\over2x}\D_\mu z^i\D^\nu\z^{\m}\D^aF_a^{\mu\nu}\]\nonumber \\ & &  
- \(5 + {f_k\f^k\over4x^2}\)\lbr\[{i\over x}\(F_a^{\nu\mu} -i\tF_a^{\nu\mu}\) 
+ {g^{\nu\mu}\over x^2}\D^a\]f_j\D_\nu z^j
K_{i\m}(T^az)^i\D_\mu\z^{\m} + {\rm h.c.}\rbr \nonumber \\ & &  
- {3f_k\f^kf_i\f_{\m}\over16x^4}\[x\(F^a_{\mu\nu} + i\tF^a_{\mu\nu}\)
\(F_a^{\mu\rho} - i\tF_a^{\mu\rho}\)\D_\rho z^i\D^\nu\z^{\m} 
+ 2i\D_\mu z^i\D^\nu\z^{\m}\D^aF_a^{\mu\nu}\] \nonumber \\ & &  
+ {\f^if_i\over2x^2}K_{i\m}\[\D\(\D_\mu z^i\D^\mu\z^{\m} + 2F^i\bF^{\m}\) 
+ i\D_\mu z^j\D_\nu\z^{\m}\D^aF_a^{\mu\nu}\] \nonumber \\ & & 
- {\f^if_i\over8x^2}\[8x\(\cW + \cbW\)\D + 
\(4x^2 - \f^jf_j\)\cW_{ab}\cbW^{ab} + 2\(\cW^{ab} + \cbW^{ab}\)\D_a\D_b\] 
\nonumber \\ & &  + {f_i\f^i\over8x^2}\[x\(F^a_{\rho\mu} 
-i\tF^a_{\rho\mu}\)\(F_a^{\rho\nu} + i\tF_a^{\rho\nu}\)\D_\nu z^i\D^\mu\z^{\m}
K_{i\m} - 4 \D^2 \] + x^4\rho_{ij}\rho^{ij}\cW\cbW \nonumber \\ & &  
+ \lbr\(F^j\bF^{\m} + \D_\mu z^j\D^\mu\z^{\m}\)\[4\rho_{\m ij}(T^az)^i\D_a -
\(\rho_{\m ij} + {\f_{\m}\over x}\rho_{ij}\)\f^i\D\] + {\rm h.c.}\rbr
\nonumber \\ & & + e^{-K/2}\lbr\cbW\[x^2\rho^{ij}\(A_{jik}F^k - 
{1\over3}A_{ij}\M\) + {f_k\f^i\f^j\over4x^2}F^kA_{ij}\] + {\rm h.c.}\rbr
\nonumber \\ & & + 2x\[4\rho_{ij}(T^az)^i(T^bz)^j\cW_{ab} 
+ i\rho_{\m ij}\D_\mu z^j\D_\nu\z^{\m}(T^az)^i\(F_a^{\mu\nu} - 
i\tF_a^{\mu\nu}\) + {\rm h.c.}\] \nonumber \\ & &
+ \lbr\rho_{ij}\D^\mu z^j\[{2\over x}\f_{\m}\D_\mu\z^{\m}(T^az)^i\D_a - 
{\f^i\over2}\cW f_j\D_\mu z^j \] +  {\rm h.c.}\rbr 
+ 2x^2\rho_{ij}\rho^j_{\m}\D\D_\rho z^i\D^\rho\z^{\m}  \nonumber \\ & &
- {i\over2}K_{i\m}\[\D^\nu\z^{\m}(T_az)^i - \D^\nu z^i(T_az)^{\m}\]
\[\f^i\rho_{ij}\D^\rho z^j\(F^a_{\rho\nu} - i\tF^a_{\rho\nu}\)+ {\rm h.c.}\] 
\nonumber \\ & & + {\D_a\over2x}\[ K_{k\m}\D^\mu\z^{\m}(T^az)^k + 
{i\over2}\f_{\m}\D_\nu\z^{\m}\(F^{\nu\mu} -i\tF^{\nu\mu}\) + {\rm h.c.}\]
\(\rho_{ij}\D_\mu z^i\f^j + {\rm h.c.}\) \nonumber \\ & & 
- \[\cW^{ab}\rho_{ij}\f^i(T_az)^j\D_b + x^2\D_\rho z^i
\D^\rho z^j\(2\rho_{ij}\cW - R_{\n i\m j}\rho^{\m\n}\cbW\) + {\rm h.c.}\],
\nonumber \\ L_g &=& {e^{-K}\over4x^2}\(\D_\mu z^i\D^\mu\z^{\m} 
+ F^i\bF^{\m}\)A_{ij}\A_{\m}^kf_k\f^j + {2\over9}M\M\(\rho_{ij}\D_\mu z^i\D^\mu
z^j + {\rm h.c.}\)  \nonumber \\ & &  
- {e^{-K/2}\over8x^3}\lbr\f^jA_{ij}F^kf_k\[\f_{\m}\(\D_\mu z^i\D^\mu\z^{\m} - 
F^i\bF^{\m}\) + \D_\mu z^if_{\ell}\D^\mu z^{\ell} - {4x\over3}F^iM\]
+ {\rm h.c.}\rbr 
\nonumber \\ & & - {e^{-K/2}\over2}\lbr \f^{\ell}A_{k\ell}\[{\f_{\m}\over3x^2}
\D^\mu\z^{\m}\D_\mu z^{\ell}\M - \rho_{ij}\D_\mu z^i\(F^j\D^\mu z^k - 
F^k\D^\mu z^j\)\] + {\rm h.c.}\rbr  \nonumber \\ & &  
+ {1\over2x^2}\lbr\D_\mu z^i\bF_i\[{x\over3}\M\(2\f_{\m}\D^\mu\z^{\m} -
f_j\D^\mu z^j\) + f_jF^j\f_{\m}\D^\mu\z^{\m}\] + {\rm h.c.}\rbr
\nonumber \\ & &  
+ {1\over4x^2}\[\D_\mu z^i\D^\mu z^jf_if_j\({1\over9}M\M + {1\over4x^2}
f_kF^k\f_{\m}\bF^{\m} + {1\over6x}f_kF^kM\) + {\rm h.c.}\]  \nonumber \\ & &  
- \lbr{f_if_j\f_{\m}\over24x^3}\[\(\D_\mu z^i\D^\mu\z^{\m} + 2F^i\bF^{\m}\)F^jM
- 2\D_\mu z^i\D^\mu z^j\bF^{\m}\M\] + {\rm h.c.}\rbr  \nonumber \\ & &  
- {1\over4x}\lbr\rho_{ij}\f^{\ell}\bF_{\ell}\D_\mu z^i\[f_k\(F^j\D^\mu z^k + 
F^k\D^\mu z^j\) + \f_{\n}F^j\D^\mu\z^{\n} 
+ {4x\over3}\D^\mu z^j\M\] + {\rm h.c.} \rbr \nonumber \\ & &  
+ {f_k\f^k\over4x}\[\cW\(x\rho_{ij}\D_\mu z^i\D^\mu z^j -
{e^{-K/2}\over2x}f_j\A^j_{\m}\bF^{\m} - {\f_{\m}f_j\over4x^2}F^j\bF^{\m} 
- {f_i\over3x}F^iM\) + {\rm h.c.}\] \nonumber \\ & &
- {1\over3}K_{i\m}K_{j\n}\(\D_\mu z^i\D^\mu\z^{\m}\D_\nu z^j\D^\nu\z^{\n}  
+ F^i\bF^{\m}F^j\bF^{\n} - F^j\bF^{\n}\D_\mu z^i\D^\mu\z^{\m}\)\nonumber \\ & & 
+ {1\over6}K_{i\m}K_{j\n}\D_\mu z^i\D_\nu\z^{\m}\(4\D^\mu z^j\D^\nu\z^{\n} +
\D^\nu z^j\D^\mu\z^{\n}\) - {1\over27}\(M\M\)^2 + {2\over9}\D M\M 
\nonumber \\ & & + \({1\over9}M\M - {1\over3}\D\)K_{i\m}\(2F^i\bF^{\m} - 
\D_\mu z^i\D^\mu\z^{\m}\) + {2\over3}\D^2 + x^2\cW_{ab}\cbW^{ab} 
\nonumber \\ & & + \(F^a_{\rho\mu} + i\tF^a_{\rho\mu}\)\(F_a^{\rho\nu} -
i\tF_a^{\rho\nu}\)\({f_i\f_{\m}\over4x} - 
{x\over2}K_{i\m}\)\D_\nu z^i\D^\mu\z^{\m} + \({f_i\f^i\over4x}\)^2\cW\cbW 
\nonumber \\ & & + {1\over2}\(\cW_{ab} + \cbW_{ab}\)\D^a\D^b 
+ x^2\rho_{ij}\rho_{\n\m}\(\D_\mu z^i\D^\mu z^j\D_\nu\z^{\m}\D^\nu\z^{n} 
+ \D_\mu z^i\D^\mu\z^{\m}F^j\bF^{\n}\) \nonumber \\ & &  
+ {f_if_j\f_{\m}\f_{\n}\over16x^4}\(\D_\mu z^i\D^\mu z^j\D_\nu\z^{\m}
\D^\nu\z^{\n} - F^iF^j\bF^{\m}\bF^{\n} - 2\D_\mu z^i\D^\mu\z^{\m}F^j\bF^{\n}\) 
\nonumber \\ & &  
+ {f_j\f_{\n}\over12x^2}F^j\bF^{\n}\[K_{i\m}\(\D_\mu z^i\D^\mu\z^{\m} -2
F^i\bF^{\m}\) + 2M\M -2\D\] \nonumber \\ & & 
+ {f_j\f_{\n}\over6x^2}\D_\nu z^j\D^\nu\z^{\n}\[K_{i\m}\(2\D_\mu z^i
\D^\mu\z^{\m} - F^i\bF^{\m}\) + {1\over2}M\M - \D\]\nonumber \\ & & 
- {f_j\f_{\m}\over2x^2}K_{j\n}\(\D_\mu z^i\D^\mu z^j
\D_\nu\z^{\m}\D^\nu\z^{\n} + \D_\mu z^i\D^\mu\z^{\n}\D_\nu\z^{\m}\D^\nu z^j\),
\eea
where indices are raised and lowered with the K\"ahler metric ($\bF_i = 
K_{i\m}\bF^{\m},\; etc.$), and 
\beq \cW^a_b = {1\over4}\(F^a_{\mu\nu}F_b^{\mu\nu} - 
iF^a_{\mu\nu}\tF_b^{\mu\nu}\) - {1\over 2x^2}\D^a\D_b, \;\;\;\; \cW= \cW^a_a
\eeq
is the bosonic part of the $F$-component of the composite chiral 
supermultiplet constructed from the Yang-Mills chiral superfield $W^a(\theta) 
= \lambda^a_L + O(\theta),$ $F^i = -e^{-K/2}\A^i$ is the bosonic part of the 
$F$-component of the chiral supermultiplet $Z^i$, $M = - 3e^{-K/2}A$ 
is an auxiliary field in the gravity supermultiplet~\cite{bggm}.  The results
of~\cite{us,us2} were calculated using the classical Lagrangian (1)
that is obtained after elimination of the auxiliary fields, and are expressed 
in those papers as functions of the boson fields and their covariant 
derivatives.  It is easy to show that calculating the one loop
corrections before or after elimination of the auxiliary fields in terms of
their classical solutions gives the same result to the loop order considered.
Here we use the auxiliary fields to present the results in a form
that lends itself more easily to an interpretation in terms of superfield 
operators.

The real function $K_R(Z,\Z)$,
\beq K_R = K + \lll\[ e^{-K}\(A_{ij}\A^{ij} -2A_i\A^i - 4A\A\) 
-4\K_a^a - \(12 + 4x^2\rho_i\rho^i\)\D\],\eeq
contains logarithmically divergent contributions
to the the renormalized K\"ahler potential.
It was shown in~\cite{mk} that the quadratically divergent term, after
Pauli-Villars regularization and an appropriate Weyl transformation, can be
absorbed into an additional renormalization of the K\"ahler potential.
In writing $L_{1,2,g}$ we have dropped total derivatives, as well as 
terms that vanish when the tree-level equations of motion are imposed, 
including a redefinition~\cite{lahanas} of the space-time metric, so as to cast
the Einstein term in canonical form.  In $L_{2,g}$ we have also
introduced scalar field reparameterization invariant covariant derivatives
($\rho_{ij},\;\rho_{\m ij}$) of the variable $\rho$, defined as the squared 
gauge coupling $\rho = x^{-1} = g^2$.

In effective supergravity from superstring theory, the classical K\"ahler
potential $K(z,\z)$, superpotential $W(z)$ and Yang-Mills normalization 
function $f_{ab}(z)$ take the forms 
\bea K(z,\z) &=& -\ln(s+\s) + G(y,\y), \;\;\;\; W(z) = W(y), \nonumber \\ 
f_{ab}(z) &=& \delta_{ab}k_as,\;\;\;\;s=z^0,\;y^i=z^{i\ge1}.\eea
In this case $1- (4x)^{-2}\f^if_i = A + (2x)^{-1}\f^iA_i
= 0$, and $\rho_i = D_i\rho = - (2x^2)^{-1}f_i $ is covariantly constant: 
$\rho_{ij} = \rho_{\m ij}= \cdots = 0.$ Then $L_2$ and $L_g$ reduce to:
\bea L_2 &=& \(\cW^{ab} + \cbW^{ab}\)\[ 3C_G\delta_{ab} - 
D_i(T_bz)^jD_j(T_az)^i\] - 24i\D_\mu z^i\D_\nu\z^{\m}K_{i\m}\D^aF_a^{\mu\nu} 
\nonumber \\ & &  
+ {N+5\over12}\[(s+\s)^2\cW_{ab}\cbW^{ab} + 2\(\cW^{ab} + \cbW^{ab}\)\D_a\D_b 
+ 8\D^2\] + {7\over2}(s+\s)^2\cW_{ab}\cbW^{ab} \nonumber \\ & & 
- {N + 5\over3}\D\[K_{i\m}\(2F^i\bF^{\m} + \D_\mu z^i\D^\mu\z^{\m}\) -
{2\over9}M\M\] + 11\(\cW^{ab} + \cbW^{ab}\)\D_a\D_b 
\nonumber \\ & & + {(s+\s)\over2}\(\cW + \cbW\)\[{2\over3}M\M + 
K_{i\m}\(\D_\rho z^i\D^\rho\z^{\m} - 2F^i\bF^{\m}\) + 2\D\] + 20\D^2 
\nonumber \\ & & 
- {N + 2\over12}(s+\s)\(F^a_{\rho\mu} - i\tF^a_{\rho\mu}\)\(F_a^{\rho\nu} + 
i\tF_a^{\rho\nu}\)\D_\nu z^i\D^\mu\z^{\m}K_{i\m} + {154\over9}M\M\D
\nonumber \\ & & +  2K_{i\m}\(13F^i\bF^{\m} + 9\D_\mu z^i\D^\mu\z^{\m}\)\D
- {2e^{-K/2}\over(s+\s)}\D_a\[(T^az)^i 
R_{i\;\;\ell}^{\;\; j\;\; k}F^{\ell}A_{jk} +{\rm h.c.}\] \nonumber \\ & &
- \[{4\over3}\D R_{i\m} + {\D_a\over(s+\s)}\(R_{\n i\m j}D^j(T^a\z)^{\n}
- {4\over3}K_{i\m}D_j(T^az)^j\)\]\(F^i\bF^{\m} + \D_\mu z^i\D^\mu\z^{\m} \) 
\nonumber \\ & & + {D_j(T^az)^j\over3}\[13iK_{\m i}F^a_{\mu\nu}\D^\mu z^i
\D^\nu\z^{\m} - {8\D_a\D\over(s+\s)}\] + 
2iF^a_{\mu\nu}D_j(T_az)^iR^j_{i\m k}\D^\mu z^k\D^\nu\z^{\m} \nonumber \\ & &
- {12\over s+\s}\lbr\[i\pp_\nu s\(F_a^{\nu\mu} - i\tF_a^{\nu\mu}\) + 
{2\pp^\mu s\over s+\s}\D_a\]\D_\mu\z^{\m}K_{i\m}(T^az)^i + {\rm h.c.}\rbr 
\nonumber \\ & & 
- 2{\pp_\rho s\pp^\nu\s\over(s+\s)}\(F^a_{\mu\nu} + i\tF^a_{\mu\nu}\)
\(F_a^{\mu\rho} - i\tF_a^{\mu\rho}\) + 40{\pp_\mu s\pp^\mu\s\over(s+\s)^2}\D + 
28i{\pp_\mu s\pp_\nu\s\over(s+\s)^2}\D^aF_a^{\mu\nu}, \nonumber \\
L_g &=& {(s+\s)^2\over4}\(\cW\cbW + \cW_{ab}\cbW^{ab}\) + {1\over2}\(\cW_{ab} + 
\cbW_{ab}\)\D^a\D^b + {2\over3}\D^2 \nonumber \\ & &  
+ {1\over6}K_{i\m}K_{j\n}\(4\D_\mu z^i\D^\mu z^j
\D_\nu\z^{\m}\D^\nu\z^{\n} + \D_\mu z^i\D^\mu\z^{\n}\D_\nu\z^{\m}\D^\nu z^j\)
\nonumber \\ & & 
- {1\over3}K_{i\m}K_{j\n}\(\D_\mu z^i\D^\mu\z^{\m}\D_\nu z^j\D^\nu\z^{\n}  
+ F^i\bF^{\m}F^j\bF^{\n} - F^j\bF^{\n}\D_\mu z^i\D^\mu\z^{\m}\)\nonumber \\ & & 
+ \({2\over27}M\M - {1\over3}\D\)K_{i\m}
\(2F^i\bF^{\m} - \D_\mu z^i\D^\mu\z^{\m}\) - {2\over27}\D M\M \nonumber \\ & & 
- {1\over81}\(M\M\)^2 - (s+\s)\(\cW + \cbW\)\({1\over2}F^i\bF^{\m}K_{i\m} - 
{1\over9}M\M\) \nonumber \\ & & 
+ {2\pp_\nu s\pp^\nu\s\over3(\s+\s)^2}\[K_{i\m}\(2\D_\mu z^i
\D^\mu\z^{\m} - F^i\bF^{\m}\) + {1\over3}M\M - \D\]\nonumber \\ & & 
+ {1\over(s+\s)^4}\pp_\mu s\pp^\mu s\pp_\nu\s\pp^\nu\s 
- {2\pp_\mu s\pp_\nu\s\over(s+\s)^2}K_{j\n}\(\D^\mu z^j
\D^\nu\z^{\n} + \D^\mu\z^{\n}\D^\nu z^j\) \nonumber \\ & & 
+ \(F^a_{\rho\mu} + i\tF^a_{\rho\mu}\)\(F_a^{\rho\nu} - i\tF_a^{\rho\nu}\)
\({\pp_\nu s\pp^\mu\s\over2(s+\s)} - 
{s+\s\over4}K_{i\m}\D_\nu z^i\D^\mu\z^{\m}\),\eea
with now:
\bea K_R &=& K + \lll\lbr e^{-K}\[A_{ij}\A^{ij} -2A_i\A^i + (N_G - 4)A\A\]
-4\K_a^a - 16\D\rbr, \nonumber \\
M &=& -{3\over s+\s}\bF^{\s},\quad M\M = 9F^s\bF_s, \eea
where the second line follows from the tree level equations of motion.

In addition, in the untwisted sector of orbifold compactifications,
the Riemann tensor is covariantly constant and its elements are related to
elements of the K\"ahler metric.  Moreover in many models there are global
symmetries that impose $R^{i\m j\n}W_{ij} = R^{i\m j\n}W_{ijk} = 0$. 
In this case $L_1$ can be expressed entirely in terms of $F^i,\D_\mu z^i,M,$ 
their complex conjugates, and the K\"ahler metric; an explicit example is 
given in~\cite{us}.

Some comments on the implications and applications of our results are in order.
It has already been shown, using the gauge fixing and expansion 
procedures defined here, that the one-loop quadratic divergences~\cite{mk}, as 
well as the logarithmic divergences~\cite{mk2} in the flat space limit and in 
the absence of a dilaton,
can be regulated \`a la Pauli-Villars.  Regularization of the full supergravity 
divergences without a dilaton are under study~\cite{rey}.  An objective of
this study is to determine the extent to which a regularization procedure can 
be achieved that preserves nonlinear symmetries of the classical effective 
Lagrangian.  To obtain the full one-loop Lagrangian, including finite 
contributions, requires a resummation of the derivative expansion.  A 
procedure for resummation will be described elsewhere~\cite{rey}.

We have presented our results for one-loop corrections to the general classical 
supergravity Lagrangian~\cite{crem,bggm} with at most two-derivative terms.
Our results for this general case can be summarized as follows.  
We define an operator of dimension $d$ as a 
K\"ahler invariant operator whose term of lowest dimension is $d$, where 
scalar and Yang-Mills fields are assigned the canonical dimension of unity.  
Then, among the ultra-violet divergent terms generated at one
loop, all operators of dimension 6 or less (as well as many operators of 
dimension 8) that involve neither the K\"ahler curvature nor derivatives of 
the gauge kinetic function can be absorbed by field redefinitions, interpreted 
as renormalizations of the K\"ahler potential, or take the form 
$F_{ab}(z,\z)\(W^aW^b\)_F + $ h.c., where $W^a$ is a chiral Yang-Mills
supermultiplet, the subscript denotes the F-component, and the matrix-valued
function $F_{ab}(z,\z)$ is not in general holomorphic.  The remaining terms of 
dimension 8 and higher must be interpreted as arising from higher order 
spinorial derivatives of superfield operators.

As seen above, the result simplifies considerably for the classical
effective Lagrangian derived from string theory, due to the the absence of a
potential for the dilaton and the special form of its K\"ahler potential.  These
features are modified when the effective Lagrangian includes a nonperturbatively
induced~\cite{nilles} superpotential for the dilaton and/or the 
Green-Schwarz counterterm~\cite{anomalies} that is necessary to restore modular
invariance.  The latter term destroys the no-scale nature of Lagrangians from
torus compactification and the untwisted sector of orbifold compactification,
and generally destabilizes the effective scalar potential.
However this term is of one-loop order and therefore should be
considered together with the full one-loop corrections.  An interesting
question, that will be addressed elsewhere, 
is whether the one-loop corrections presented here can restabilize the 
potential.

An important unresolved issue in the construction of effective supergravity
Lagrangians for gaugino condensation is the correct form of the kinetic term
for the composite chiral multiplet that represents the lightest bound state of
the confined Yang-Mills sector.  It has recently been shown~\cite{kin}, in the 
context of both the linear and chiral multiplet formulations for the dilaton,
that such terms can be generated by operators of higher dimension.  The
effective Lagrangian (17) determines the leading one-loop
contribution to the relevant operators; similar operators occur in string
theory~\cite{tomd}. Thus the determination of
loop corrections can provide a guide to the construction of the
effective theory, which in turn could shed light on
gaugino condensation as a mechanism for supersymmetry breaking.

\vskip .3in
\noindent{\bf Acknowledgements.} This work was supported in part by the
Director, Office of Energy Research, Office of High Energy and Nuclear Physics,
Division of High Energy Physics of the U.S. Department of Energy under Contract
DE-AC03-76SF00098 and in part by the National Science Foundation under grant
PHY-95-14797 and PHY-93-09888.


\begin{thebibliography}{99}
\bibitem{us} M.K. Gaillard and V. Jain, {\it Phys. Rev.} {\bf D49:} 1951 
(1994); misprints and errors in this paper are corrected in~\cite{us2}.
\bibitem{us2} M.K.~Gaillard, V.~Jain and K. Saririan, LBL-34948, UCB-PTH-93/37,
ITP-SB-95-38 (1995) hep-th/9606052.
\bibitem{crem} E. Cremmer, S. Ferrara, L. Girardello, and A. Van 
Proeyen, {\it Nucl. Phys.} {\bf B212:} 413 (1983).
\bibitem{bggm} P. Bin\'etruy, G. Girardi, R. Grimm and M. Muller, 
{\it Phys. Lett.} {\bf 189B:} 83 (1987); P. Bin\'etruy, G. Girardi and R. 
Grimm, Annecy preprint LAPP-TH-275-90, (1990).
\bibitem{barb} R. Barbieri, S. Ferrara, L. Maiani, F. Palumbo and C.A. Savoy,
{\it Phys. Lett.} {\bf 115B:} 212 (1982).
\bibitem{ant} I. Antoniadis, E. Gava, K.S. Narain and T.R. Taylor, 
{\it Nucl. Phys.} {\bf B407:} 706 (1993).
\bibitem{tini} G.K. 't Hooft and M. Veltman, {\it Ann. Inst. Henri Poincar\'e}
{\bf 20:} 69 (1974).
\bibitem{noncan} M.K. Gaillard and V. Jain, {\it Phys. Rev.} {\bf D46:} 1786 
(1992).
\bibitem{ino} R. Barbieri and C. Cecotti, {\it Z. Phys.} {\bf C17:} 183 
(1983); P. Bin\'etruy and M.K. Gaillard, {\it Nucl. Phys.} {\bf  B254:} 388
(1985).\bibitem{josh} J. W. Burton, M.K. Gaillard and V. Jain, {\it Phys. Rev.} 
{\bf D41:} 3118 (1990).
\bibitem{lahanas} C. Chiou-Lahanas, A. Kapella-Economu, A.B. Lahanas and 
X.N. Maintas, {\it Phys. Rev.} {\bf D42:} 469 (1990) and {\it Phys. Rev.} 
{\bf D45:} 534 (1992).
\bibitem{mk} M.K. Gaillard, {\it Phys.Lett.} {\bf B342:} 125 (1995). 
\bibitem{russ} M.A. Shifman, {\it Nucl. Phys.} {\bf B352:} 87 (1991);
M.A. Shifman, and A.I. Vainshtein {\it Nucl. Phys.} {\bf B365:} 312 (1991);
A.A. Iogansen, {\it Sov. J. Nucl. Phys.} {\bf 54} (1991).
\bibitem{frad} E.S. Fradkin and A.A. Tseytlin, {\it Ann. Phys.} {\bf 162:}
31 (1985). This paper showed equivalence up to finite topological anomalies;
subsequently full equivalence has been shown; S.J. Rey, private communication.
\bibitem{mk2} M.K. Gaillard, {\it Phys. Lett} {\bf B347:} 284 (1995).
\bibitem{rey} M.K. Gaillard and S.-J. Rey, in progress.
\bibitem{nilles} H.P. Nilles, {\it Phys.\ Lett.} {\bf 115B:} 455 (1982).
\bibitem{anomalies} G.L. Cardoso and B.A. Ovrut, {\it Nucl. Phys.} 
{\bf B369:} 351 (1992); J.-P. Derendinger, S. Ferrara, C. Kounnas and F. 
Zwirner, {\it Phys. Lett.} {\bf B271:} 307 (1991); P. Bin\'etruy, 
G. Girardi, R. Grimm and M. M\"uller, {\it Phys. Lett.} {\bf B265} 111 (1991);
P. Adamietz, P. Bin\'etruy, G. Girardi and R. Grimm, {\it Nucl. Phys.} 
{\bf B401:} 257 (1993); P. Mayr and S. Stieberger, 
{\it Nucl. Phys.} {\bf B412:} 502 (1994); 
M.K. Gaillard and T.R. Taylor, {\it Nucl. Phys.} {\bf B381:} 577 (1992);
V. Kaplunovsky and J. Louis, {\it Nucl. Phys.} {\bf B422:} 57 (1994).
\bibitem{kin} P. Bin\'etruy, M.K. Gaillard and T.R. Taylor, {\it Nucl. Phys.} 
{\bf B455:} 97 (1995);
P. Bin\'etruy and M.K. Gaillard, {\it Phys. Lett.} {\bf B365:} 87 (1996).
\bibitem{tomd} I. Antoniadis, E. Gava, K.S. Narain and T.R. Taylor,
{\it Nucl. Phys.} {\bf 428:} 282 (1994).

\end{thebibliography}
\end{document}